\documentclass[aps,prd,preprint,floats,nofootinbib]{revtex4-1}

\usepackage{amsmath,amssymb,bbm}
\usepackage{epsfig}
\usepackage{array}
\usepackage{comment}
\usepackage[pdftex,colorlinks=false,urlcolor=blue,linkcolor=blue]{hyperref}
\usepackage{epstopdf}
\newcommand{\PreserveBackslash}[1]{\let\temp=\\#1\let\\=\temp}
\newcolumntype{C}[1]{>{\PreserveBackslash\centering}p{#1}}
\newcolumntype{R}[1]{>{\PreserveBackslash\raggedleft}p{#1}}
\newcolumntype{L}[1]{>{\PreserveBackslash\raggedright}p{#1}}

\def\tanb{\tan\beta}

\def\sbma{\sin(\beta-\alpha)}
\def\cbma{\cos(\beta-\alpha)}
\def\gam{\gamma}

\def\mh{m_h}
\def\mH{m_H}

\def\ha{A}
\def\mha{m_\ha}

\def\hl{h}
\def\hh{H}
\def\mhl{m_{\hl}}
\def\mhh{m_{\hh}}

\def\CU{C_U}
\def\CD{C_D}

\def\CV{C_V}

\def\cu{\CU}
\def\cd{\CD}
\def\cv{\CV}

\def\gev{~{\rm GeV}}

\def\br{{\rm BR}}

\def\eg{{\it e.g.}}
\def\ie{{\it i.e.}}

\def\beq{\begin{equation}}
\def\eeq{\end{equation}}
\def\bea{\begin{eqnarray}}
\def\eea{\end{eqnarray}}
\def\bit{\begin{itemize}}
\def\eit{\end{itemize}}
\def\ben{\begin{enumerate}}
\def\een{\end{enumerate}}
\def\rgghgamgam{\mu_{gg}^h(\gam\gam)}
\def\rggHgamgam{\mu_{gg}^H(\gam\gam)}
\def\rvbfhgamgam{\mu_{\rm VBF}^h(\gam\gam)}
\def\rvbfHgamgam{\mu_{\rm VBF}^H(\gam\gam)}

\def\rgghzz{\mu_{gg}^h(ZZ)}
\def\rggHzz{\mu_{gg}^H(ZZ)}

\def\br{{\mathcal B}}

\def\brnew{\br_{\rm new}}

\def\lsim{\mathrel{\raise.3ex\hbox{$<$\kern-.75em\lower1ex\hbox{$\sim$}}}}
\def\gsim{\mathrel{\raise.3ex\hbox{$>$\kern-.75em\lower1ex\hbox{$\sim$}}}}

\begin{document}
\title{Addendum to ``Constraints on and future prospects for Two-Higgs-Doublet Models in light of the LHC Higgs signal''}

\author{B\'eranger Dumont$^{1}$}
\email[]{dumont@lpsc.in2p3.fr}
\author{John F.~Gunion$^2$}
\email[]{jfgunion@ucdavis.edu}
\author{Yun~Jiang$^2$}
\email[]{yunjiang@ucdavis.edu}
\author{Sabine Kraml$^{1}$}
\email[]{sabine.kraml@lpsc.in2p3.fr}

\affiliation{(1) \,Laboratoire de Physique Subatomique et de Cosmologie, Universit\'e Grenoble-Alpes,
CNRS/IN2P3, 53 Avenue des Martyrs, F-38026 Grenoble, France}
\affiliation{(2) \,Department of Physics, University of California, Davis, CA 95616, USA}

\begin{abstract}
We update the constraints on Two-Higgs-Doublet Models of Type~I and II discussed in 
arXiv:1405.3584  
using the latest LHC measurements of the $\sim 125.5\gev$ Higgs signal as of Summer 2014.  
We provide explicit comparisons of the results before and after the Summer 2014 ATLAS and CMS updates.  
Overall, the changes with respect to arXiv:1405.3584 are rather small;  to a large extent this is due to the fact that both the ATLAS  and the CMS updates of the $\gamma\gamma$ decay mode moved closer to SM expectations.  
\end{abstract}


\maketitle

\section{Introduction}

In a recent paper \cite{Dumont:2014wha}, we provided a comprehensive analysis of the status of 
Two-Higgs-Doublet Models (2HDMs) of Type I and Type II, considering both the cases where the 
observed Higgs particle at the LHC is the lighter CP-even state $h$ or the heavier CP-even state $H$.
To this end, we performed scans of the 2HDM parameter space taking into account all relevant 
constraints from precision electroweak data, from stability, unitarity and perturbativity of the potential, 
as well as from $B$ physics and from the direct searches at LEP. 
We also employed the most recent limits from searches for heavy Higgs-like states at the LHC. 
The central piece of the analysis however was to check for  consistency with the various 
signal strength measurements of the observed $\sim 125.5\gev$ Higgs boson at the LHC, 
including a consistent treatment of ``feed down'' from the production of heavier
Higgs states. This was done based on the results published by ATLAS and CMS by June 2013, 
\ie\ according to the status of the Moriond and LHCP 2013 conferences 
(see \cite{Belanger:2013xza} for a summary).

Since then, a number of new measurements or updates of existing ones were published by the experimental collaborations. Most significant, from the point of view of our analysis of the 2HDMs, were the 
long-awaited final results for the $\gamma\gamma$ decay mode from CMS~\cite{Khachatryan:2014ira} 
in July and the update of the $\gamma\gamma$ results from ATLAS~\cite{Aad:2014eha} at the end of August 2014. 
There were also several other important new measurements or updates; for example uncertainties have been 
significantly reduced for the fermionic channels, particularly for $H \to b\bar b$ in ttH production. 
All these new results were put together and analyzed in global coupling fits in~\cite{Bernon:2014vta}. 

In the present note, we now revisit the analysis of \cite{Dumont:2014wha} and 
ask what are the implications in the 2HDM context of all these new (or updated) results on the signal strengths of the 
$\sim 125.5\gev$ Higgs boson. To this aim, we take the scan points of \cite{Dumont:2014wha} and, 
leaving everything else the same, update the $\chi^2$ calculation for the signal strengths 
at $125.5$~GeV in the $\mu_{\rm ggF+ttH}(Y)$ versus $\mu_{\rm VBF+VH}(Y)$ planes, 
Eq.~(5) of \cite{Dumont:2014wha}, with the new numbers presented in Table~I of~\cite{Bernon:2014vta}.
Points for which $\chi^2_Y<6.18$ for each decay mode $Y=\gamma\gamma, VV(=WW,ZZ), b\bar b, \tau\tau$ (that means points that are consistent within 95.4\% confidence level (CL)  
with the observed signal strengths for each decay mode $Y$) and that in addition pass 
all other relevant constraints  
will be called ``postLHC8(2014)-FDOK'' and 
compared to the corresponding points of~\cite{Dumont:2014wha}, called  ``postLHC8(2013)-FDOK''. 
In the plots, we will moreover identify the points that fit both the 2013 and the 2014 analyses  
as ``postLHC8(2013 \& 2014)-FDOK''. 

We note that in this addendum we focus on pointing out the small modifications that arise from the latest ATLAS and CMS results. For a detailed physics discussion and implications for future measurements, \eg\ at the next run of the LHC at $\sqrt{s}=13-14$~TeV, we refer to the main paper~\cite{Dumont:2014wha}, whose results overall remain perfectly valid. 
For notations and conventions, we also refer to~\cite{Dumont:2014wha}.

\section{\bf \boldmath $\mh\sim 125.5\gev$ scenarios}
\label{hlsection}

\begin{figure}[b!]
\begin{center}
\includegraphics[width=0.49\textwidth]{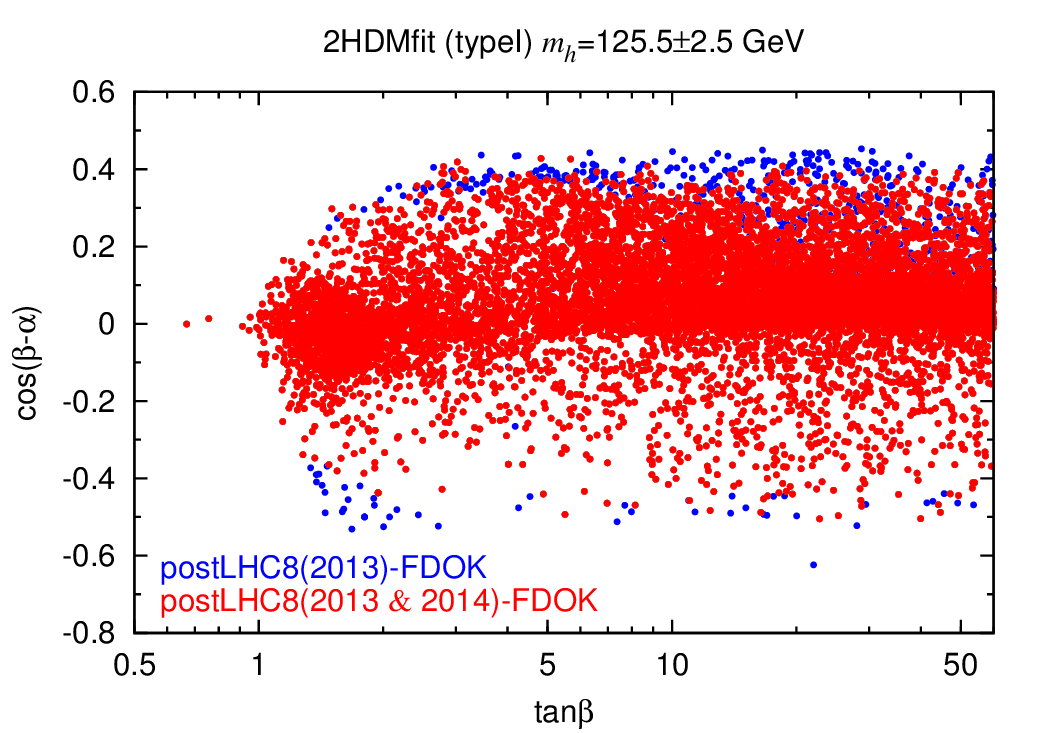}
\includegraphics[width=0.49\textwidth]{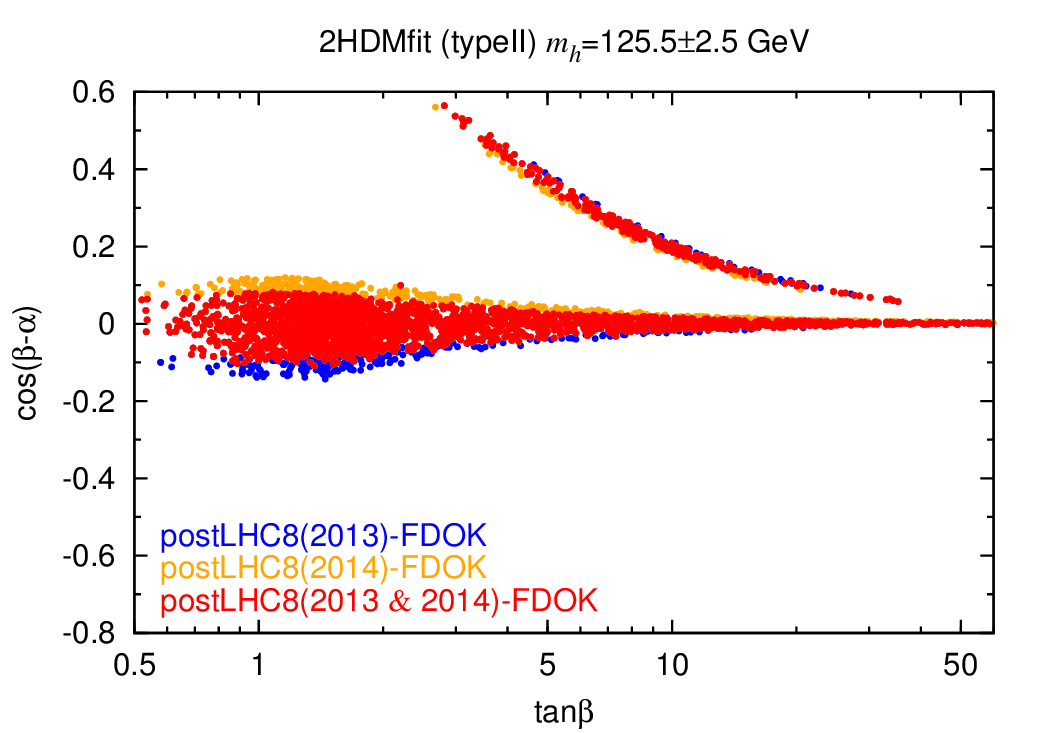}
\end{center}\vspace*{-5mm}
\caption{Constraints in the $\cbma$ versus $\tanb$ plane for $\mh\sim 125.5\gev$.  Blue points are those which passed all constraints given the Higgs signal strengths as of Spring 2013 (\ie\ the status considered in \cite{Dumont:2014wha}), 
red points are those which remain valid when employing the Summer 2014 updates, and  
orange points are those newly allowed after the Summer 2014 updates.  
Note that the latter occur only in Type~II but not in Type~I models.}
\label{fig:h1-cosba-tb}
\end{figure}

We begin by showing in Fig.~\ref{fig:h1-cosba-tb} the points surviving all constraints in the $\cbma$ versus $\tanb$ plane for the $\mh\sim 125.5\gev$ scenarios. 
The red points satisfy the Higgs signal constraints from both 2013 and 2014, while orange (blue) points pass only the 2014 (2013) Higgs constraints.   
We see that, in Type~I, large values of $|\!\cbma|$ get slightly more constrained, while in Type~II there is a narrow strip around $\cbma\approx -0.1$ and $\tan\beta\lesssim 2$ that is now excluded. On the other hand, in Type~II slightly larger positive values of $\cbma$ are allowed from the 2014 measurements. (Such orange points, which were not compatible with the 2013 results but are now allowed after the 2014 updates do occur only in Type~II but not in Type~I.) 
The banana-shaped branch spanning from $(\tan\beta,\cbma)\approx(3,0.6)$ to $(40,0.1)$ is still present; this corresponds to the solution with a flipped sign for $\cd$.

\begin{figure}[t!]
\begin{center}
\includegraphics[width=0.49\textwidth]{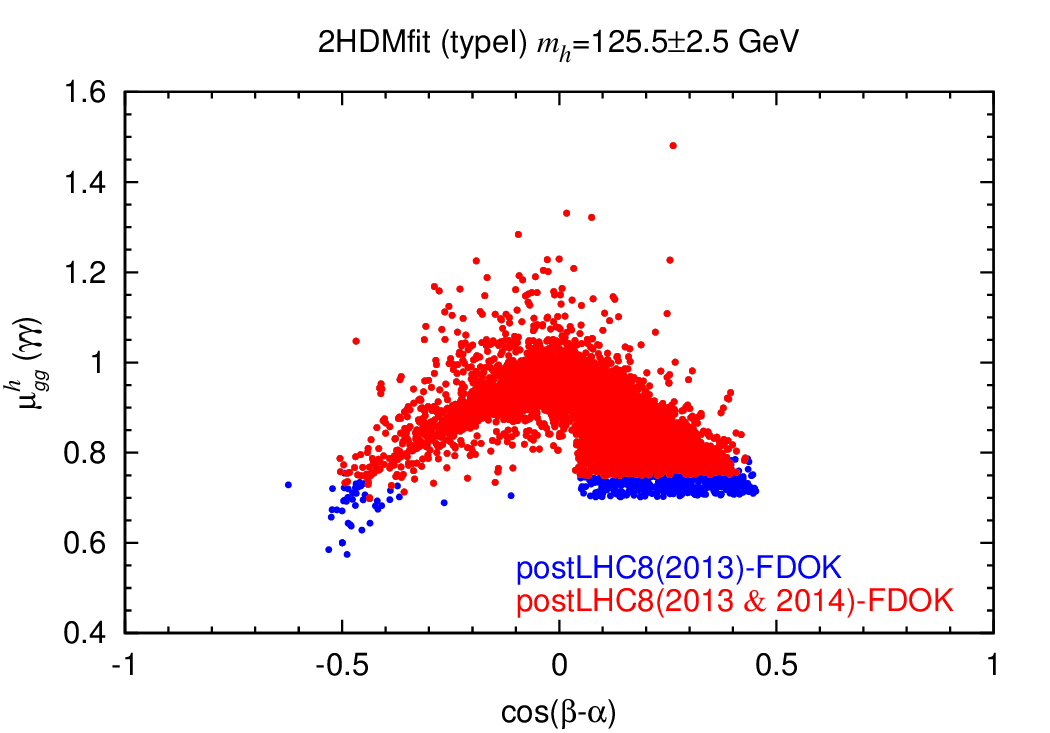}
\includegraphics[width=0.49\textwidth]{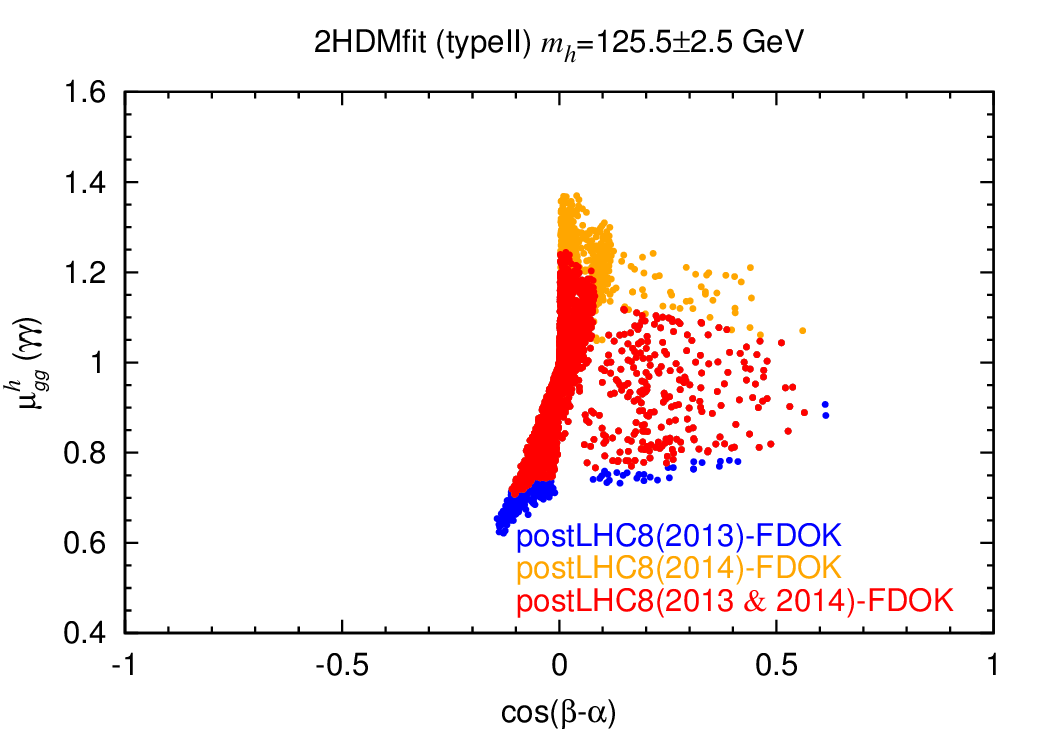}
\includegraphics[width=0.49\textwidth]{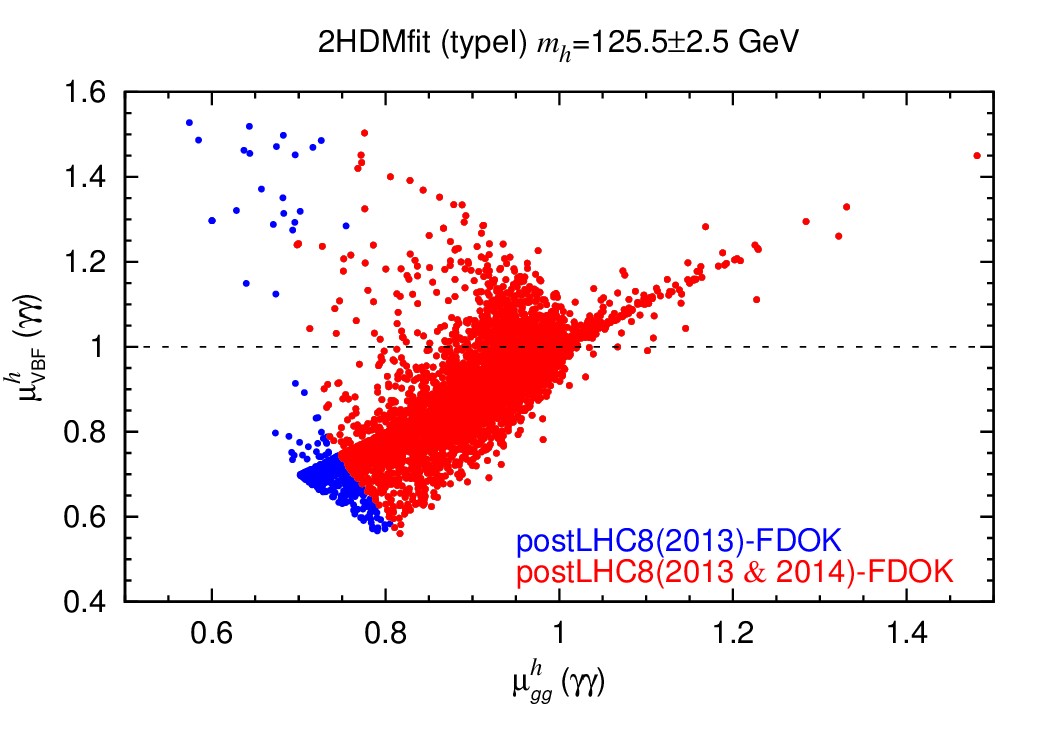}
\includegraphics[width=0.49\textwidth]{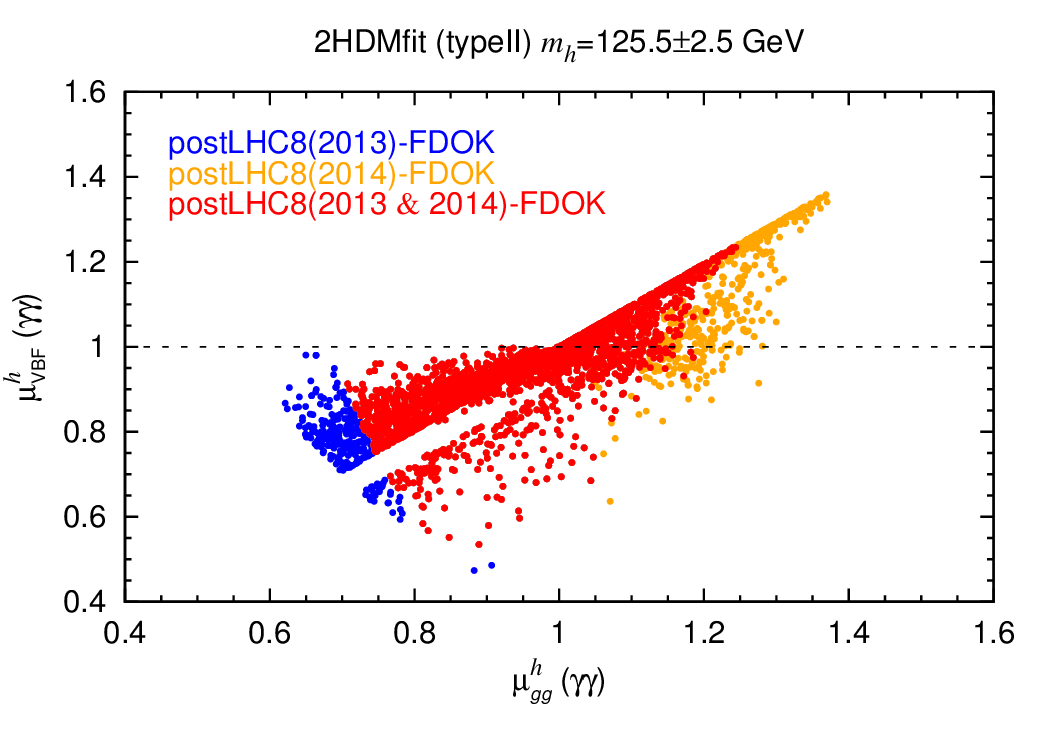}
\includegraphics[width=0.49\textwidth]{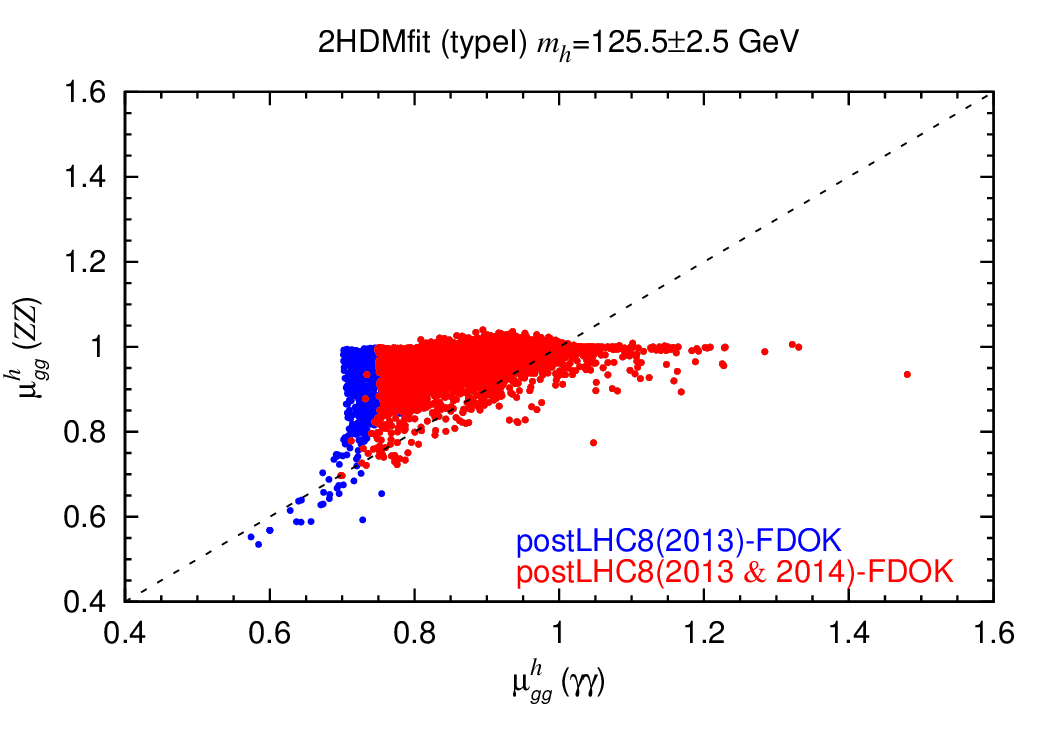}
\includegraphics[width=0.49\textwidth]{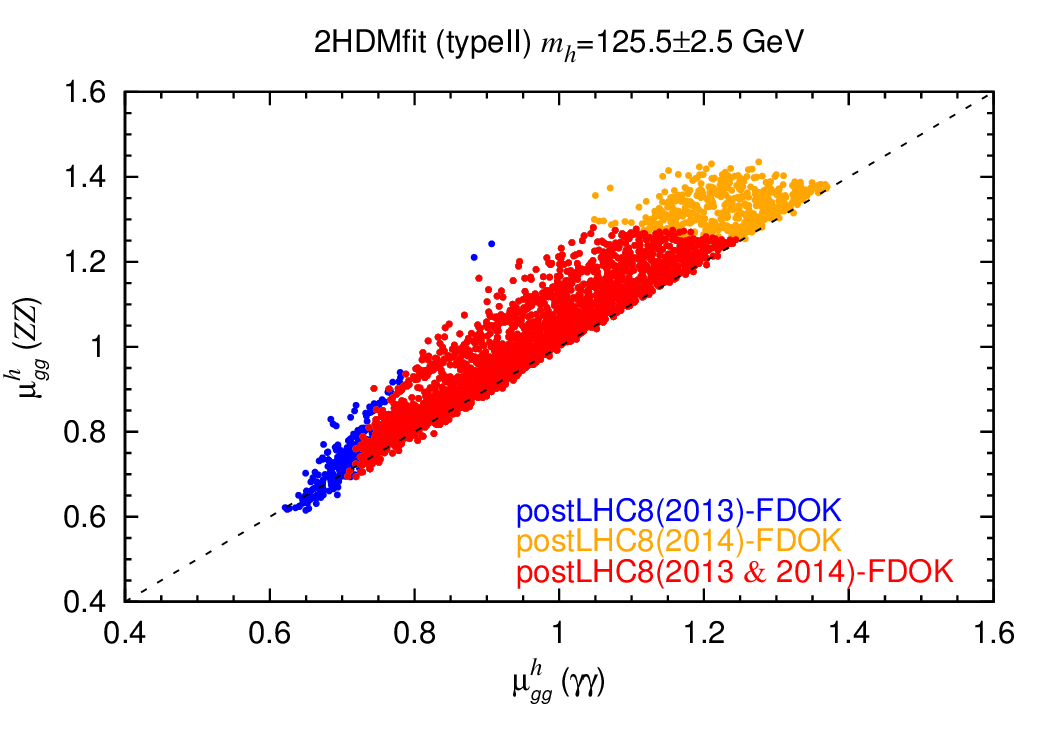}\\
\end{center}\vspace*{-5mm}
\caption{As in Fig.~\ref{fig:h1-cosba-tb}, but for $\rgghgamgam$ and $\rvbfhgamgam$ signal strengths versus $\cbma$ (top row), $\rvbfhgamgam$ versus\ $\rgghgamgam$ (middle row), and $\rgghzz$  versus\ $\rgghgamgam$ (bottom row).}
\label{fig:h1-mu-cosba}
\end{figure}

The reason for these slight changes lies mostly in the new combined signal strengths 
for the $\gamma\gamma$ decay mode: 
$\widehat\mu_{\rm{ggF+ttH}}({\gamma\gamma})=1.25\pm 0.24$ and 
$\widehat\mu_{\rm{VBF+VH}}({\gamma\gamma})=1.09\pm0.46$ 
with a correlation of $\rho=-0.30$~\cite{Bernon:2014vta}, as compared to 
$\widehat\mu_{\rm{ggF+ttH}}({\gamma\gamma})=0.98\pm 0.28$ and 
$\widehat\mu_{\rm{VBF+VH}}({\gamma\gamma})=1.72\pm0.59$ 
with a correlation of $\rho=-0.38$ in 2013~\cite{Belanger:2013xza}.
The result, after combining ATLAS and CMS data, is that the best-fit signal strength in the ggF mode\footnote{In ttH production, uncertainties are still too large to have any impact.} has increased (although the new central value is consistent at the $1\sigma$ level with the 2013 results) while that in VBF+VH production has come down by a bit more than $1\sigma$. 
The $\gamma\gamma$ signal strengths as a function of $\cbma$ are shown in the top row of Fig.~\ref{fig:h1-mu-cosba}. 
Here, one sees explicitly  that the low values of $\rgghgamgam\approx 0.6-0.7$ are no longer allowed. 
Moreover, while there has been no change in the maximum $\rgghgamgam$ obtainable in Type~I, higher values
up to about $\rgghgamgam\approx1.4$ (vs. $\sim 1.2$ before) are attainable in Type~II. 
The alert reader will have noticed that this upper limit is much lower than the $2\sigma$ range 
that should be allowed in principle. 
The limitation in fact comes from the $h\to VV(=WW,ZZ)$ decay mode in ggF production, 
for which we have $\widehat\mu_{\rm{ggF+ttH}}({VV})=1.03\pm 0.17$ from the 2014 measurements, and hence $\mu^h_{gg}(ZZ)<1.37$ at 95.4\% CL  (as compared to $\widehat\mu_{\rm{ggF+ttH}}({VV})=0.91\pm 0.16$ in Spring 2013). 
The correlations between signal strengths in different channels are illustrated in the middle and bottom 
rows of Fig.~\ref{fig:h1-mu-cosba}.  

\begin{figure}[t]
\begin{center}
\includegraphics[width=0.49\textwidth]{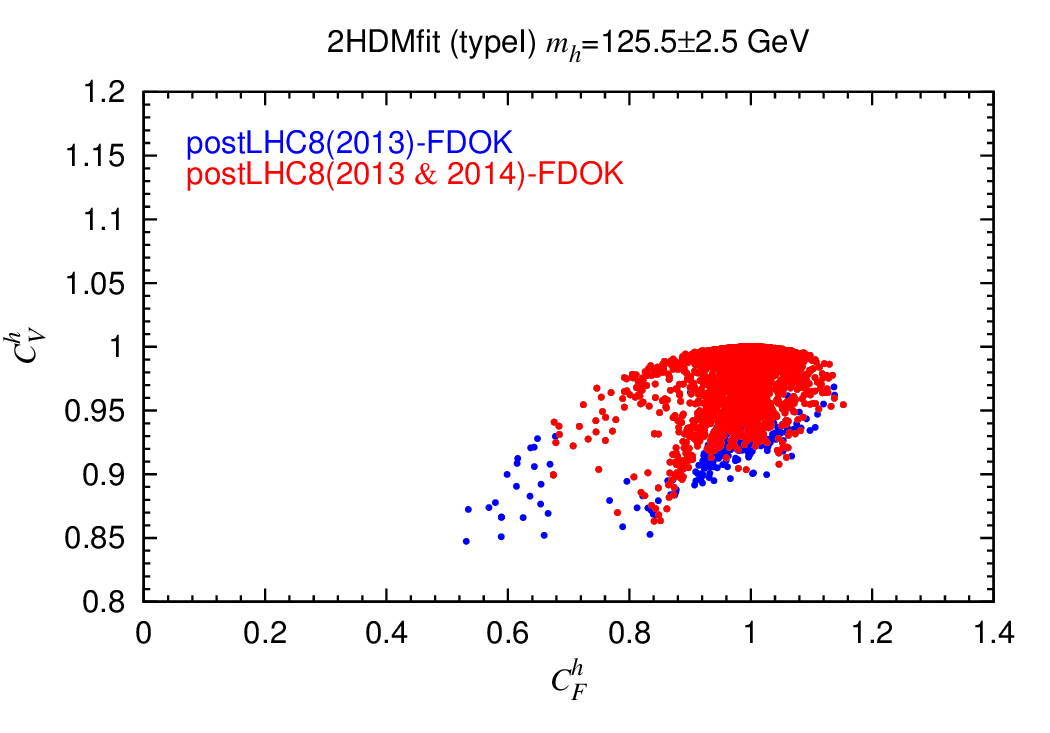}
\includegraphics[width=0.49\textwidth]{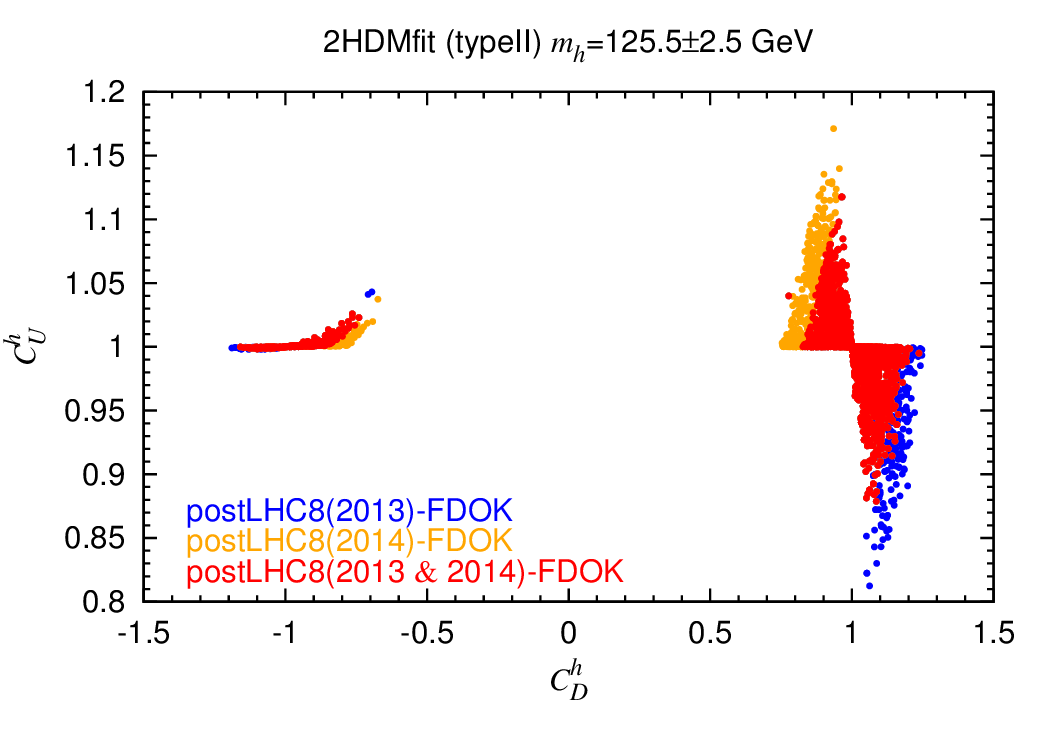}
\end{center}\vspace*{-5mm}
\caption{Reduced couplings $\cv^h$ vs.\ $C_F^h$ in Type~I (left plot) and $\cu^h$ vs.\ $\cd^h$ in Type~II (right plot) for $\mh\sim 125.5\gev$ scenarios.}
\label{fig:h1-couplings}
\end{figure}

In Fig.~\ref{fig:h1-couplings}, we plot $\cv^h$ vs.\ $C_F^h$ (Type~I) and $\cu^h$ vs.\ $C_D^h$ (Type~II) for the $\mh\sim 125.5\gev$ scenarios. Substantial suppression of the fermionic couplings is possible in both models.  
In both cases, an enhanced $h\to \gamma\gamma$ rate can come from a suppression of $\cd$, which suppresses $\br (h\to b\bar b)$. In Type~I, since $\cu=\cd\equiv C_F$, this goes hand-in-hand with a reduction of the  
$hgg$ coupling; depending on which effect dominates, $\rvbfhgamgam$ can be enhanced, while $\rgghgamgam$ is suppressed (cf. the middle-left plot in Fig.~\ref{fig:h1-mu-cosba}). This does not occur in Type~II, where enhancement/suppression of $\cu$ and $\cd$ is anti-correlated. In this case $\cd<1$ leads to an enhancement and $\cd>1$ to a suppression of both $\rgghgamgam$ and $\rvbfhgamgam$; since $\cu$ works in the same direction, 
the effect can be more pronounced for $\rgghgamgam$ than for $\rvbfhgamgam$ 
(cf. the middle-right plot in Fig.~\ref{fig:h1-mu-cosba})

\begin{figure}[t]
\begin{center}
\includegraphics[width=0.49\textwidth]{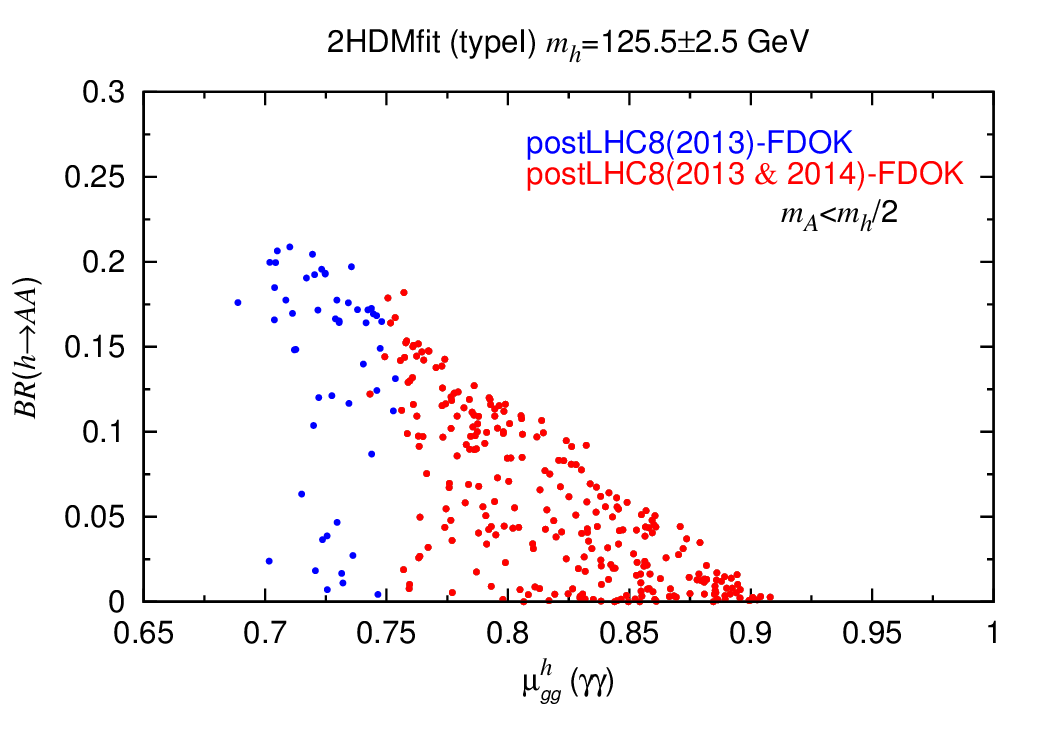}
\includegraphics[width=0.49\textwidth]{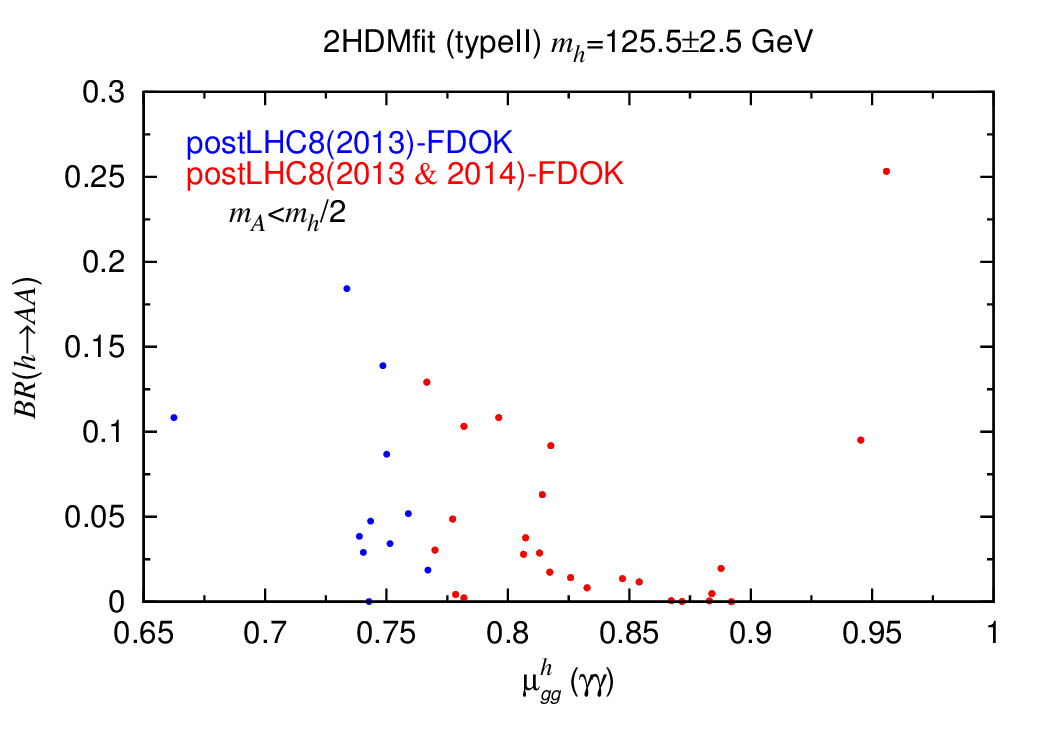}
\end{center}\vspace*{-5mm}
\caption{Correlation of BR$(h\to AA)$ and $\rgghgamgam$ for $\mh\sim 125.5\gev$ scenarios with $m_A<m_h/2$.}
\label{fig:h1-brhaa}
\end{figure}

Regarding the allowed ranges of $m_H$, $m_A$ and $m_{H^\pm}$, there is no visible change 
with respect to~\cite{Dumont:2014wha}.  
It is however interesting to take a closer look at the region $m_A<m_h/2$, where  $h\to AA$ decays are possible. 
In Fig.~\ref{fig:h1-brhaa}, we show the correlation between $\br(h\to AA)$ and $\rgghgamgam$. 
As one can see, large values of the former imply suppression of the latter. 
Although the effect is small, the requirement $\mu^h_{gg}(\gamma\gamma)>0.77$ at $2\sigma$  
constrains the maximum $\br(h\to AA)$ that can be obtained in Type~I and Type~II models. (In Type~II, however, a focused scan would be needed for a quantitative interpretation of the results.)
This limit is actually stronger than the ``direct'' constraint on unseen decays, $\brnew<0.22$~\cite{Bernon:2014vta}, 
from the generic $\cu,\ \cd,\ \cv<1$ coupling fit.

\section{\bf \boldmath $\mH\sim 125.5\gev$ scenarios}
\label{hhsection}

\begin{figure} [t!]
\begin{center}
\includegraphics[width=0.49\textwidth]{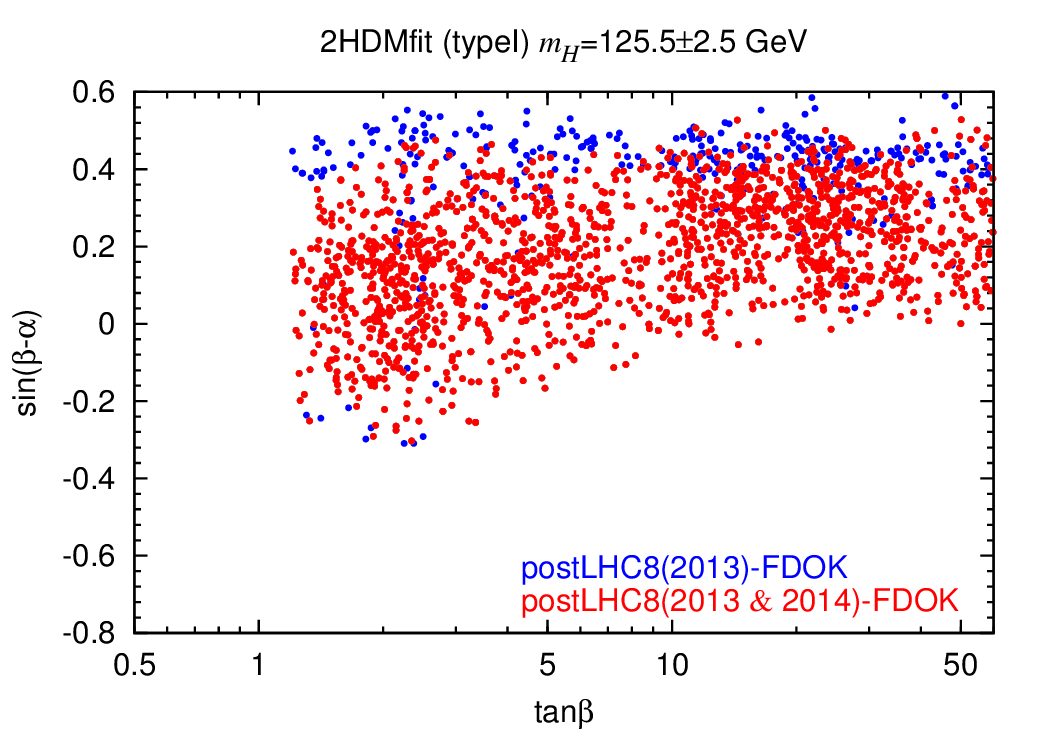}
\includegraphics[width=0.49\textwidth]{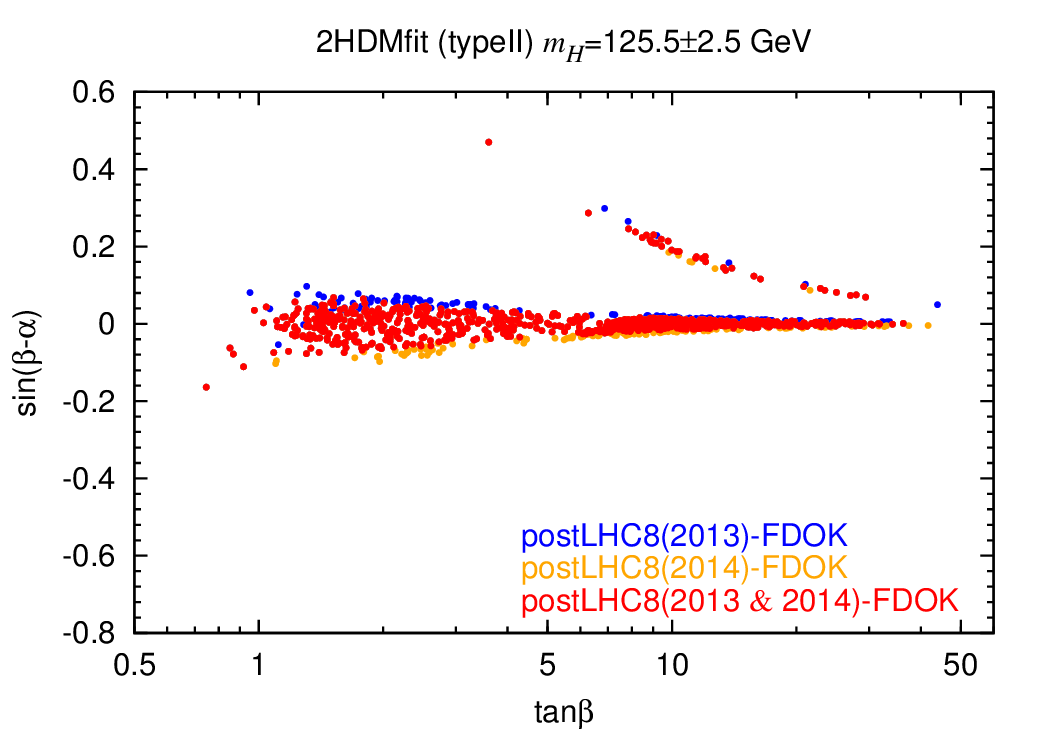}
\end{center}\vspace*{-5mm}
\caption{Constraints in the $\sbma$ versus $\tanb$ plane for $\mH\sim 125.5\gev$, comparing the current status 
as of Summer 2014  to that of Spring 2013.}
\label{fig:h2-sinba-tb}
\end{figure}

\begin{figure} [t!]
\begin{center}
\includegraphics[width=0.49\textwidth]{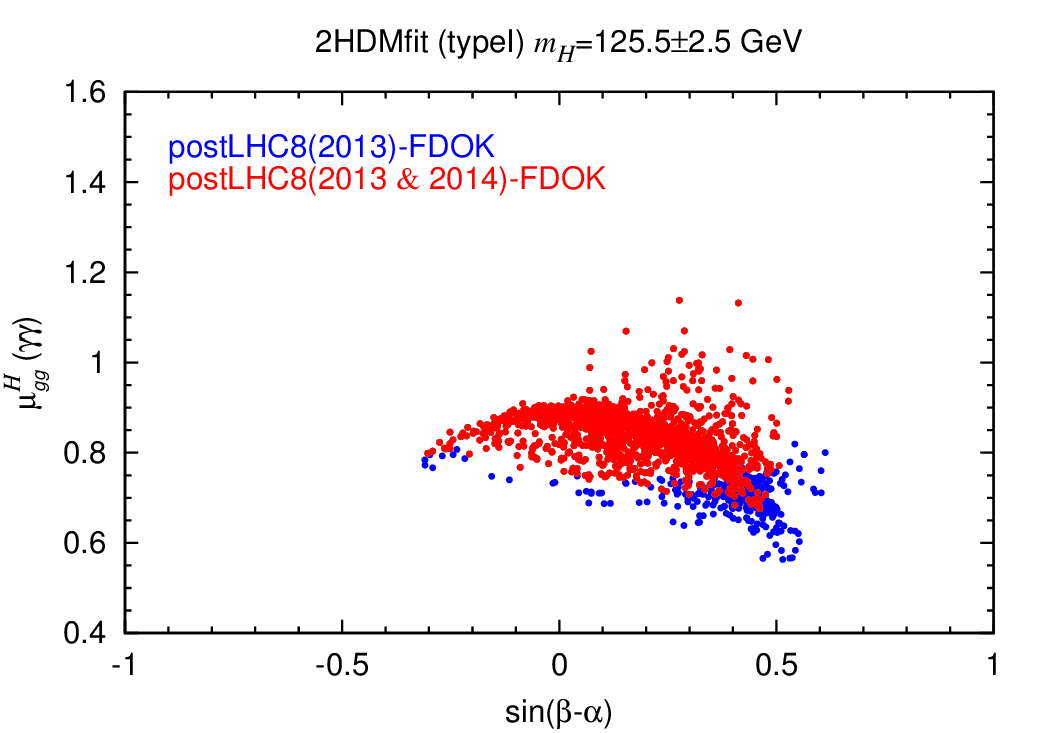}
\includegraphics[width=0.49\textwidth]{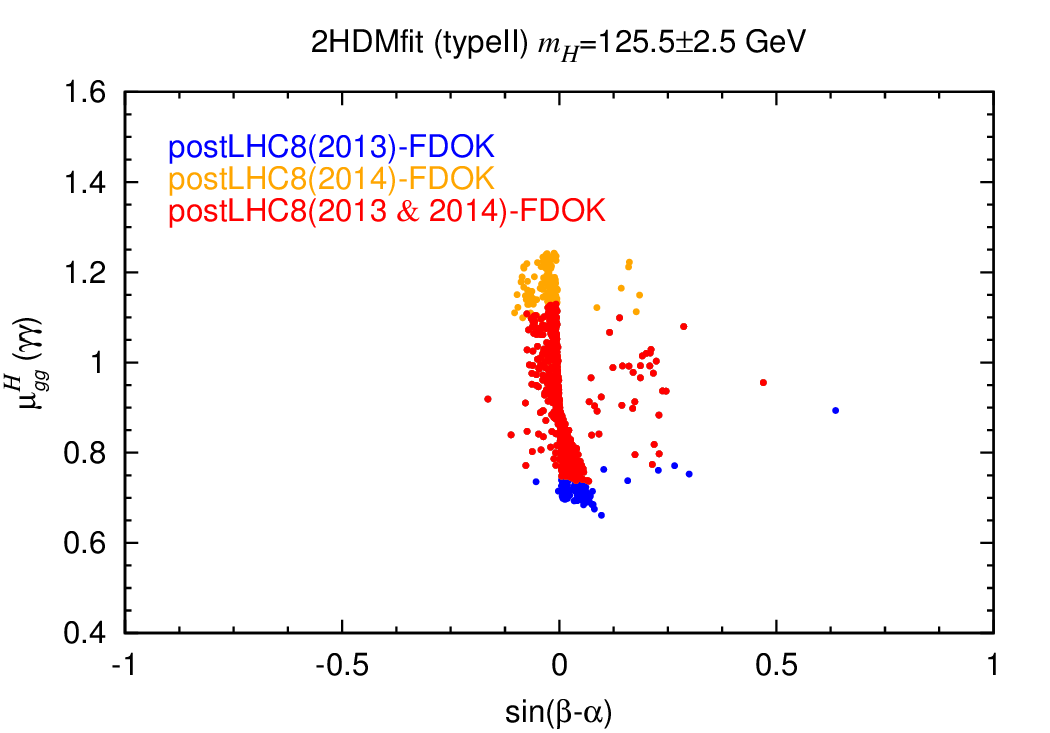}
\includegraphics[width=0.49\textwidth]{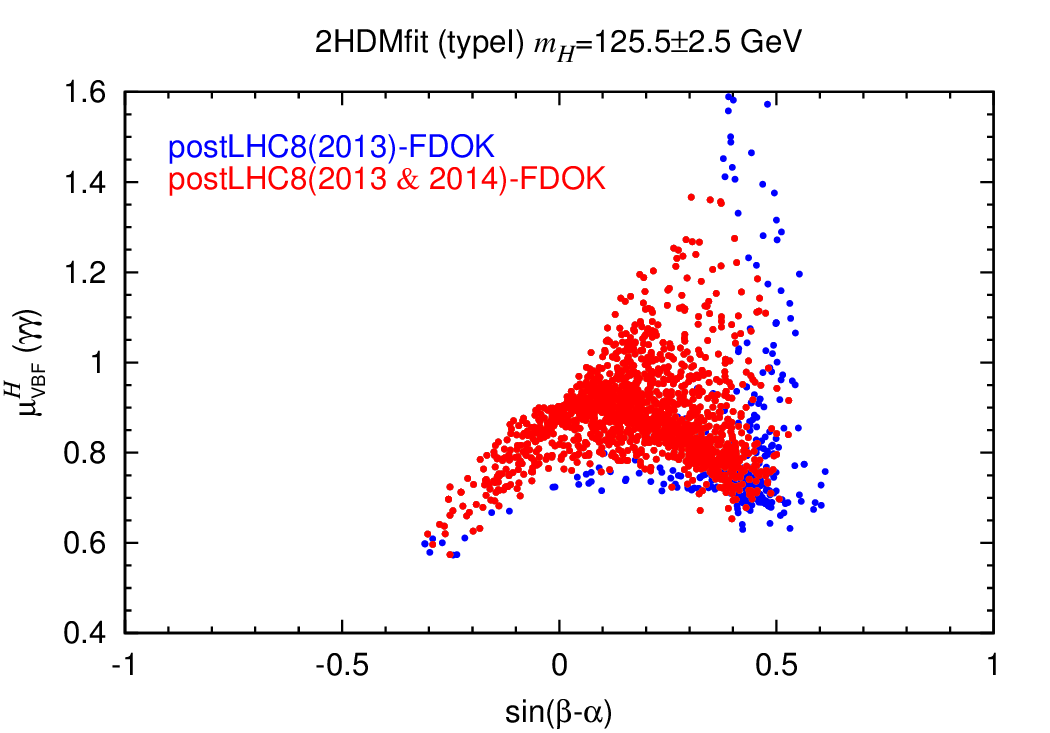}
\includegraphics[width=0.49\textwidth]{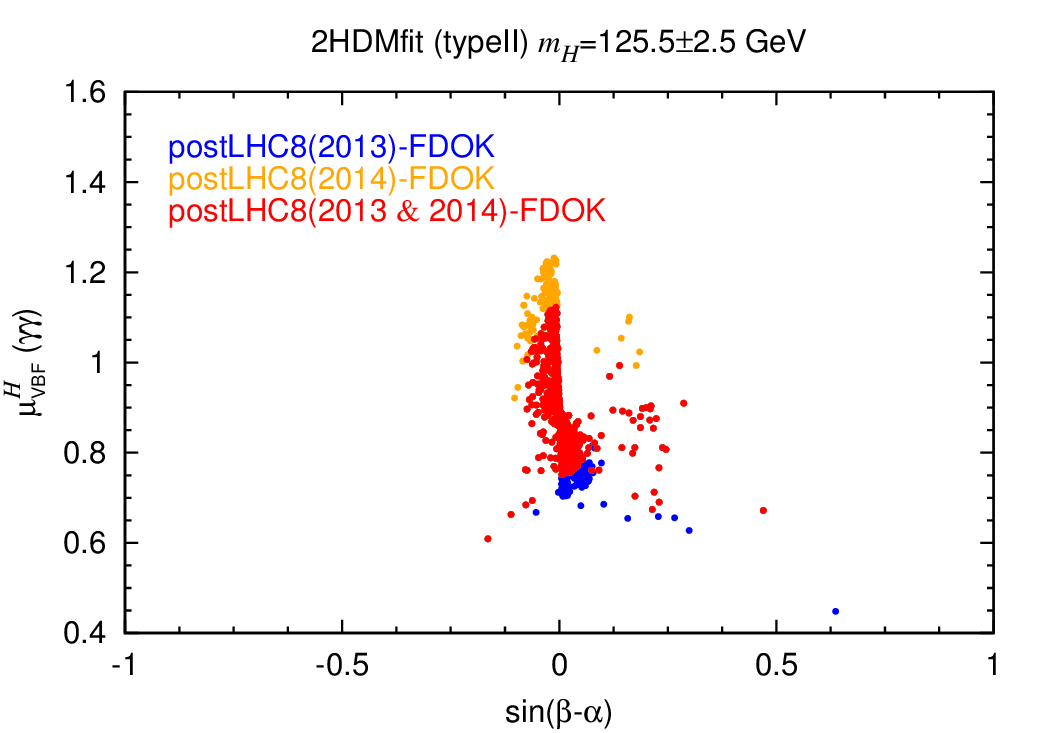}
\end{center}\vspace*{-5mm}
\caption{As Fig.~\ref{fig:h2-sinba-tb}, but for the $\gam\gam$ signal strengths versus $\sin(\beta-\alpha)$.}
\label{fig:h2-mu-sinba}
\end{figure}

Let us now turn to the case that the observed Higgs state at $\sim125.5$~GeV is the heavier CP-even scalar $H$. 
Analogous to Fig.~\ref{fig:h1-cosba-tb}, we show in Fig.~\ref{fig:h2-sinba-tb} the $\mH\sim 125.5\gev$ 
points in the $\sin(\beta-\alpha)$ versus $\tanb$ plane after all constraints have been applied. 
As before, we observe a slight narrowing of the allowed $\sin(\beta-\alpha)$ range, but no visible change in the $\tan\beta$ direction. It is interesting to note, however, that in Type~I $\mH\sim 125.5\gev$ requires $\tan\beta\gtrsim 1$. 
It is also remarkable that, while in Type~I $\sin(\beta-\alpha)$ can still vary from about $-0.3$ to $0.5$ (corresponding to $\cv\gtrsim 0.87$), in Type~II one is pretty much forced into the decoupling/alignment regime, the few points with sizable $\sin(\beta-\alpha)$ being quite rare and associated with the branch having a negative sign for $C_D^H$.

\begin{figure}[t!]
\begin{center}
\includegraphics[width=0.49\textwidth]{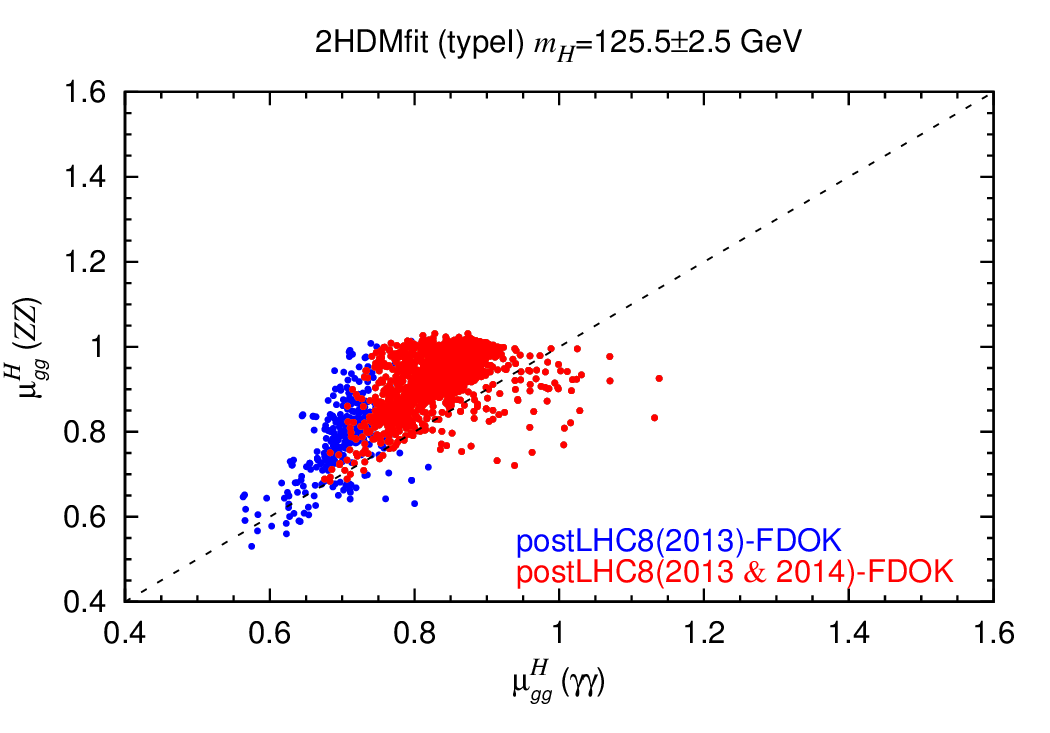}
\includegraphics[width=0.49\textwidth]{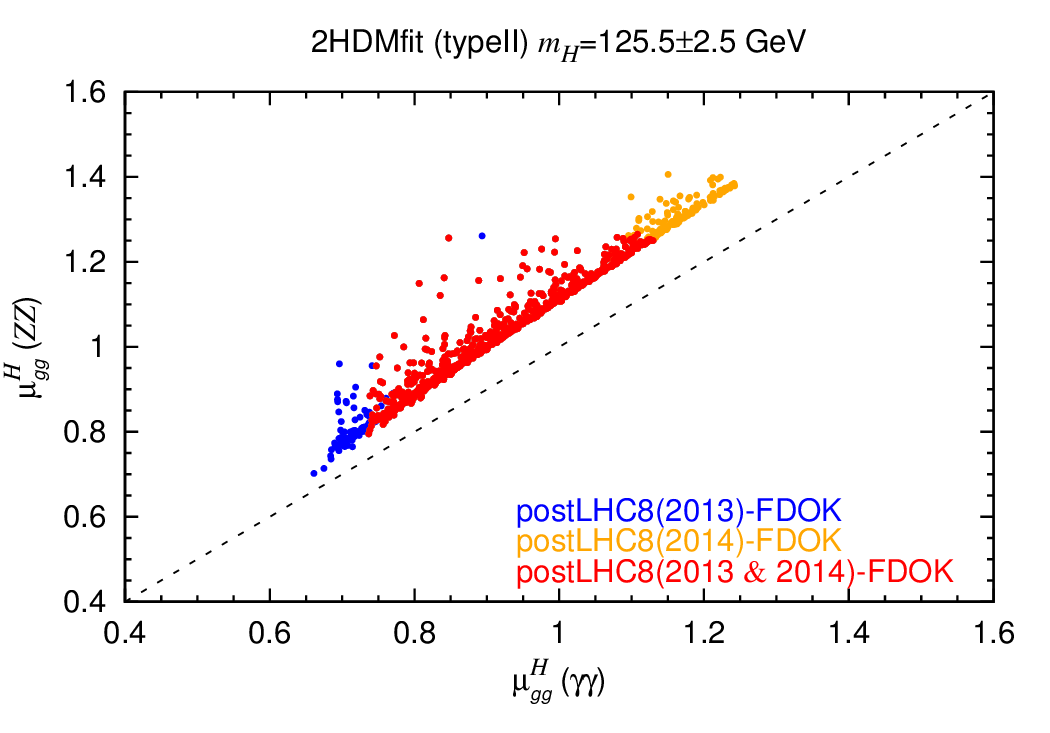}\\
\includegraphics[width=0.49\textwidth]{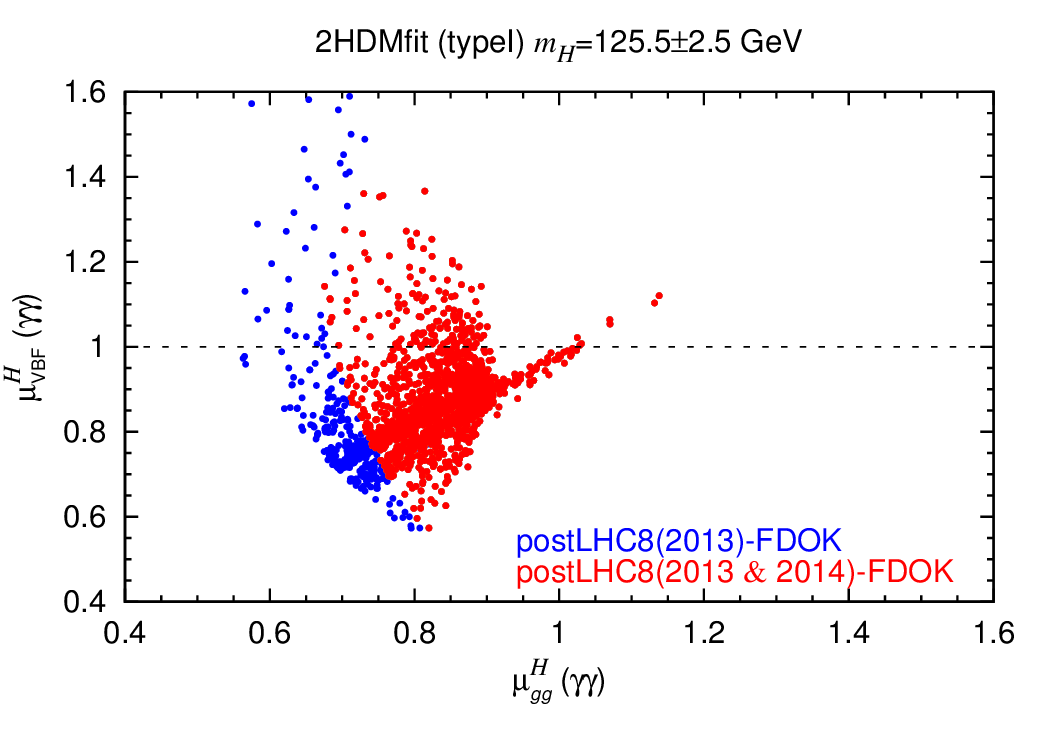}
\includegraphics[width=0.49\textwidth]{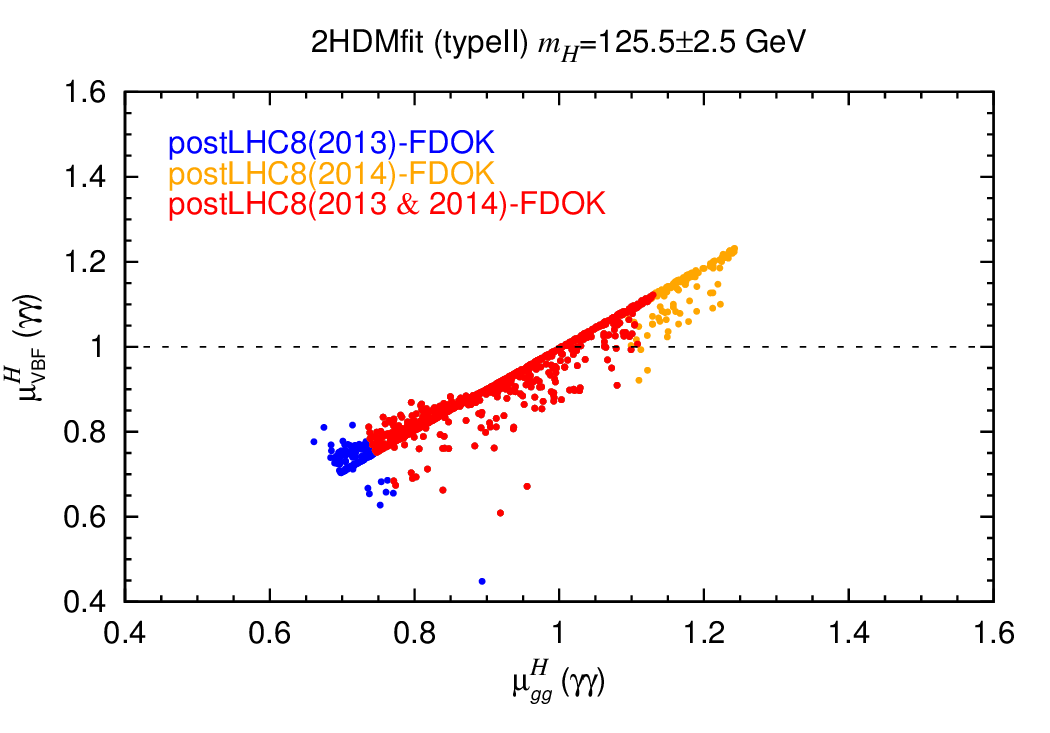}
\end{center}\vspace*{-5mm}
\caption{Correlations of signal strengths for $\mH\sim 125.5\gev$ scenarios.} 
\label{fig:h2-mu-mu}
\end{figure}

The $\gam\gam$ signal strengths as a function of $\sin(\beta-\alpha)$ are shown in Fig.~\ref{fig:h2-mu-sinba}. 
Correlations of signal strengths are illustrated in Fig.~\ref{fig:h2-mu-mu}. Analogous arguments as for the 
$\mh\sim 125.5\gev$ case apply. It is however worth noting that the direct correlation between 
$\rggHgamgam$, $\rvbfHgamgam$ and $\rggHzz$ in Type~II is much stronger than for $\mh\sim 125.5\gev$. 
As above, additional Type~II points occur with $\rggHgamgam$ and $\rggHzz$ values beyond those found in \cite{Dumont:2014wha}.  These would be removed if  future measurements show that $\rggHgamgam$ ($\rggHzz$) is within $10\%$ ($20\%$) of unity. As was noted in~\cite{Dumont:2014wha},  if $\leq \pm 5\%$ deviations from the SM are required for both the $ZZ$ and $\gam\gam$ final states then the upper plots show that only a few points  of the Type~I model having $\mu_{gg}^\hh(\gam\gam)\gsim 0.95$ can survive and that {\it all} Type~II points will be removed by this constraint. 

As regards the $h$, $A$ and even the $H^\pm$ masses associated with a good fit by the $H$ to the LHC data and other limits, there is not much change with respect to  \cite{Dumont:2014wha}. In particular the range of $h$ and $A$ masses discussed in \cite{Dumont:2014wha} remains valid, the only modification is the slight narrowing in $\sbma$  already visible in Fig.~\ref{fig:h2-sinba-tb}.

\section{Conclusions}\label{conclusions}

Overall, the new ATLAS and CMS analyses from Summer 2014 lead to relatively minor modifications 
of the preferred parameter ranges in 2HDM models of Type~I and Type~II, the most significant changes 
being slight upward shifts of the central $\mu_{gg}(\gam\gam)$ and $\mu_{gg}(VV)$ values. 
In both Type~I and Type~II this results in the exclusion of points with too low $\mu_{gg}(\gam\gam)$ 
and/or $\mu_{gg}(VV)$. In addition, in Type~II  points with somewhat higher $\mu_{gg}(\gam\gam)$ 
and $\mu_{gg}(VV)$  (beyond those allowed in the 2013 analysis) are now allowed; such new points 
however do  not occur in Type~I.
Apart from these small shifts the results and in particular the conclusions of~\cite{Dumont:2014wha} do not change.

A possibly important particular point is that  the scenarios with low $\mha<100\gev$ that escape all LEP and (so far) LHC limits and yet have quite substantial $gg\to A$ and $bbA$ production cross sections survive the latest data. 
It will be interesting to probe these scenarios, which are possible for both Type~I and Type~II in the $\mhl\sim 125.5\gev$ case and for Type~I in the $\mhh\sim 125.5\gev$ case, in ongoing analyses of the 8 TeV LHC data and in future LHC running at higher energy.

\section{Acknowledgements}

We thank J\'er\'emy Bernon for discussions on the new signal strength constraints. 
This work was supported in part by US DOE grant DE-SC-000999 and by the 
``Investissements d'avenir, Labex ENIGMASS''.  
Y.J.\ is supported by  LHC-TI fellowship  US NSF grant PHY-0969510.

\bibliographystyle{apsrev4-1}
\bibliography{2hdmfit_v2}

\end{document}